\documentclass{aa} 
\usepackage{txfonts}
\usepackage{epsfig,graphicx,aalongtable}
\usepackage{natbib}
\usepackage{times}

\bibpunct{(}{)}{;}{a}{}{,}

\begin{document}


\title{Spectroscopic metallicities for planet-host stars: extending 
       the samples\thanks{Based on observations collected 
       at La Silla Observatory, ESO, Chile, with the FEROS spectrograph
       at the 2.2-m MPI/ESO telescope (programs 72.C-0033 and 74.C-0135) and
       the CORALIE spectrograph at the 1.2-m Euler Swiss Telescope, with the
       UVES spectrograph at the VLT 8.2-m Kueyen telescope (programs 074.C-0134), and
       with the SARG spectrograph at the TNG telescope, operated at the island of 
       La Palma.}}


\author{N.C.~Santos\inst{1,2} \and
        G.~Israelian\inst{3} \and
	M.~Mayor\inst{2} \and
	J.P.~Bento\inst{1} \and
	P.C.~Almeida\inst{1} \and
	S.G.~Sousa\inst{1} \and
	A.~Ecuvillon\inst{3}
	}

\offprints{Nuno C. Santos, \email{Nuno.Santos@oal.ul.pt}}

\institute{
        Centro de Astronomia e Astrof{\'\i}sica da Universidade de Lisboa,
        Observat\'orio Astron\'omico de Lisboa, Tapada da Ajuda, 1349-018
        Lisboa, Portugal
     \and
	Observatoire de Gen\`eve, 51 ch.  des 
	Maillettes, CH--1290 Sauverny, Switzerland
     \and
        Instituto de Astrof{\'\i}sica de Canarias, E-38200
        La Laguna, Tenerife, Spain
        }
	
\date{Received / Accepted } 

\titlerunning{Spectroscopic metallicities for planet-host stars}


\abstract{
We present stellar parameters and metallicities for 
29 planet-host stars, as well as for a large volume-limited sample of 
53 stars not known to be orbited by any planetary-mass companion. 
These stars add to the results presented in our previous series
of papers, providing two large and uniform samples of 
119 planet-hosts and 94 ``single'' stars with accurate stellar parameters 
and [Fe/H] estimates. The analysis of the results further confirms that stars
with planets are metal-rich when compared with average field dwarfs.
Important biases that may compromise future studies are also discussed.
Finally, we compare the metallicity distributions for single planet-hosts
and planet-hosts in multiple stellar systems. The results show that
a small difference cannot be excluded, in the sense that the latter sample
is slighly overmetallic. However, more data are needed to confirm this correlation. 
}

\maketitle

\section{Introduction}

The discovery that stars with giant planet companions are particularly metal-rich
when compared to single field dwarfs is providing crucial clues to our
understanding of the processes of planetary formation and evolution.
This fact, first discussed by \citet[][]{Gonzalez-1997} and \citet{Gonzalez-1998},
was later proved by the uniform analysis of large samples of planet-host 
and field ``single'' stars \citep[][]{Santos-2001}.

Further refinements to these studies have shown that the probability of
finding (and presumably forming) a planet, at least of the kind now discovered
by radial-velocity surveys, is a strong function of the stellar metal content 
\citep[][]{Santos-2001,Santos-2003,Santos-2004b,Reid-2002,Fischer-2005}. Although the exact dependence is
still not completely settled \citep[see][and references therein]{Santos-2004b},
it was shown that about 25\% of solar type stars with twice the solar metallicity
are orbited by a planet, while less than 5\% of stars with the metal content 
of the Sun have an orbiting planetary-mass companion.

\begin{table*}
\caption[]{Planet-host stars studied in this paper. See text for more details.}
\begin{tabular}{lcccrccccc}
\hline
HD     & T$_{\mathrm{eff}}$ & $\log{g}_{spec}$ & $\xi_{\mathrm{t}}$ & \multicolumn{1}{c}{[Fe/H]} & N(\ion{Fe}{i},\ion{Fe}{ii}) & $\sigma$(\ion{Fe}{i},\ion{Fe}{ii}) & Instr.$\dagger$ & Mass         & $\log{g}_{hipp}$ \\
number & [K]                &  [cm\,s$^{-2}$]  &  [km\,s$^{-1}$]  &        &                              &                                    &            & [M$_{\sun}$] & [cm\,s$^{-2}$] \\
\hline
\object{BD$-$10\,3166} & 5325$\pm$45  & 4.36$\pm$0.21 & 0.95$\pm$0.05 & 0.35$\pm$0.05 & 35,9  & 0.05,0.10 & [a]  &  --  &   --\\
\object{HD\,2638}   & 5192$\pm$38  & 4.29$\pm$0.25 & 0.86$\pm$0.04 & 0.16$\pm$0.04    & 35,6  & 0.04,0.12 & [g]  & 0.82 & 4.50\\
\object{HD\,27894}  & 4875$\pm$81  & 4.22$\pm$0.26 & 0.65$\pm$0.14 & 0.30$\pm$0.07    & 35,5  & 0.07,0.10 & [g]  & 0.68	& 4.43\\
\object{HD\,37605}  & 5391$\pm$49  & 4.37$\pm$0.18 & 0.97$\pm$0.06 & 0.31$\pm$0.06    & 38,6  & 0.05,0.08 & [a]  & 0.95 & 4.54\\
\object{HD\,41004\,A}  & 5242$\pm$57  & 4.35$\pm$0.22 & 1.01$\pm$0.07 & 0.16$\pm$0.07 & 39,6  & 0.06,0.10 & [a]  & 0.80 & 4.39\\
\object{HD\,59686}  & 4871$\pm$135 & 3.15$\pm$0.41 & 1.85$\pm$0.12 & 0.28$\pm$0.18    & 39,10 & 0.16,0.18 & [a]  & --   &   --\\
\object{HD\,63454}  & 4841$\pm$65  & 4.23$\pm$0.30 & 0.89$\pm$0.09 & 0.11$\pm$0.07    & 35,5  & 0.06,0.15 & [g]  & 0.69 & 4.57\\
\object{HD\,70642}  & 5671$\pm$46  & 4.39$\pm$0.27 & 1.01$\pm$0.06 & 0.20$\pm$0.06    & 39,12 & 0.05,0.11 & [a]  & 0.98 & 4.42\\
\object{HD\,70642}  & 5693$\pm$26  & 4.41$\pm$0.09 & 1.01$\pm$0.04 & 0.18$\pm$0.04    & 36,8  & 0.03,0.04 & [c]  & 0.99 & 4.43\\
\object{HD\,70642}  & 5682         & 4.40          & 1.01          & 0.19             & --    & --        & avg. & 0.98 & 4.42\\
\object{HD\,83443}  & 5501$\pm$63  & 4.46$\pm$0.36 & 1.07$\pm$0.08 & 0.39$\pm$0.07    & 39,11 & 0.07,0.16 & [a]  & 0.96 & 4.41\\
\object{HD\,83443}  & 5454$\pm$61  & 4.33$\pm$0.17 & 1.08$\pm$0.08 & 0.35$\pm$0.08    & 38,7  & 0.07,0.08 & [c]  & 0.93 & 4.37\\
\object{HD\,83443}  & 5478         & 4.40          & 1.08          & 0.37             & --    & --        & avg. & 0.94 & 4.39\\
\object{HD\,88133}  & 5438$\pm$34  & 3.94$\pm$0.11 & 1.16$\pm$0.03 & 0.33$\pm$0.05    & 35,7  & 0.04,0.04 & [d]  & 0.98 & 3.83\\
\object{HD\,92788}  & 5758$\pm$37  & 4.30$\pm$0.16 & 1.10$\pm$0.04 & 0.34$\pm$0.05    & 38,11 & 0.04,0.07 & [a]  & 1.06 & 4.44\\
\object{HD\,92788}  & 5821$\pm$41  & 4.45$\pm$0.06 & 1.16$\pm$0.05 & 0.32$\pm$0.05    & 37,5  & 0.04,0.02 & [c]  & 1.12 & 4.49\\
\object{HD\,92788}  & 5790         & 4.38          & 1.13          & 0.33             & --    & --        & avg. & 1.09 & 4.46\\
\object{HD\,93083}  & 4995$\pm$50  & 4.26$\pm$0.19 & 0.66$\pm$0.08 & 0.15$\pm$0.06    & 36,5  & 0.05,0.08 & [h]  & 0.70 & 4.43\\
\object{HD\,99492}  & 4810$\pm$72  & 4.21$\pm$0.21 & 0.72$\pm$0.13 & 0.26$\pm$0.07    & 33,4  & 0.07,0.09 & [d]  & --   & --\\
\object{HD\,101930} & 5079$\pm$62  & 4.24$\pm$0.16 & 0.73$\pm$0.08 & 0.17$\pm$0.06    & 36,5  & 0.05,0.06 & [h]  & 0.74	& 4.40\\
\object{HD\,102117} & 5708$\pm$46  & 4.34$\pm$0.10 & 1.15$\pm$0.06 & 0.33$\pm$0.06    & 39,8  & 0.05,0.03 & [a]  & 1.05 & 4.25\\
\object{HD\,102117} & 5672$\pm$22  & 4.27$\pm$0.07 & 1.05$\pm$0.02 & 0.30$\pm$0.03    & 37,8  & 0.03,0.03 & [h]  & 1.03 & 4.23\\
\object{HD\,102117} & 5690         & 4.30          & 1.10          & 0.32             & --    & --        & avg. & 1.04	& 4.24\\
\object{HD\,104985} & 4773$\pm$62  & 2.76$\pm$0.14 & 1.71$\pm$0.07 & $-$0.28$\pm$0.09 & 33,9  & 0.09,0.05 & [b]  &  --  &   --\\
\object{HD\,106252} & 5834$\pm$37  & 4.22$\pm$0.12 & 1.06$\pm$0.06 & $-$0.03$\pm$0.05 & 39,11 & 0.04,0.06 & [a]  & 0.98 & 4.35\\
\object{HD\,106252} & 5899$\pm$35  & 4.34$\pm$0.07 & 1.08$\pm$0.06 & $-$0.01$\pm$0.05 & 37,6  & 0.04,0.04 & [c]  & 1.02 & 4.39\\
\object{HD\,106252} & 5866         & 4.28          & 1.07          & $-$0.02          & --	 & --        & avg. & 1.00 & 4.37\\
\object{HD\,114386} & 4865$\pm$93  & 4.30$\pm$0.30 & 0.86$\pm$0.12 & $-$0.04$\pm$0.07 & 37,8  & 0.06,0.13 & [a]  & 0.66 & 4.52\\
\object{HD\,114386} & 4804$\pm$61  & 4.36$\pm$0.28 & 0.57$\pm$0.12 & $-$0.08$\pm$0.06 & 35,4  & 0.06,0.14 & [c]  & 0.54 & 4.40\\
\object{HD\,114386} & 4834         & 4.33          & 0.72          & $-$0.06          & --	 & --        & avg. & 0.60 & 4.46\\
\object{HD\,117207} & 5654$\pm$33 & 4.32$\pm$0.05 & 1.13$\pm$0.04 & 0.23$\pm$0.05     & 37,7  & 0.04,0.02 & [d]  & 0.99  & 4.33\\
\object{HD\,117618} & 6013$\pm$41 & 4.39$\pm$0.07 & 1.73$\pm$0.09 & 0.06$\pm$0.06     & 37,7  & 0.05,0.03 & [d]  & 1.10  & 4.35\\
\object{HD\,142022\,A} & 5499$\pm$27 & 4.36$\pm$0.04 & 0.94$\pm$0.03 & 0.19$\pm$0.04  & 36,6  & 0.03,0.03 & [i]  & 0.90  & 4.33\\
\object{HD\,147513}  & 5894$\pm$31   & 4.43$\pm$0.17 & 1.26$\pm$0.05 &  0.08$\pm$0.04 & 39,11  & 0.04,0.07 & [a] & 1.13  & 4.54\\
\object{HD\,147513}  & 5883$\pm$25   & 4.51$\pm$0.05 & 1.18$\pm$0.04 &  0.06$\pm$0.04 & 36,7   & 0.03,0.03 & [c] & 1.11  & 4.53\\
\object{HD\,147513}  & 5888          & 4.47          & 1.22          &  0.07          & --     & --        & avg.& 1.12  & 4.54\\
\object{HD\,154857} & 5610$\pm$27  & 4.02$\pm$0.12 & 1.30$\pm$0.05 & $-$0.23$\pm$0.04 & 38,8  & 0.03,0.05 & [a]  & 1.29 & 3.79\\
\object{HD\,160691} & 5813$\pm$42  & 4.25$\pm$0.07 & 1.30$\pm$0.05 & 0.32$\pm$0.05    & 37,8  & 0.04,0.03 & [e]  & 1.10 & 4.25\\
\object{HD\,160691} & 5798$\pm$33  & 4.31$\pm$0.08 & 1.19$\pm$0.04 & 0.32$\pm$0.04    & 36,7  & 0.04,0.03 & [c]  & 1.10 & 4.25\\
\object{HD\,160691} & 5806         & 4.28          & 1.24          & 0.32             & --    & --        & avg. & 1.10 & 4.25\\
\object{HD\,183263} & 5991$\pm$28  & 4.38$\pm$0.17 & 1.23$\pm$0.04 & 0.34$\pm$0.04    & 38,8  & 0.03,0.07 & [a]  & 1.18 & 4.36\\
\object{HD\,188015} & 5793$\pm$39  & 4.49$\pm$0.10 & 1.14$\pm$0.05 & 0.30$\pm$0.05    & 38,7  & 0.04,0.04 & [a]  & 1.08 & 4.41\\
\object{HD\,208487} & 6141$\pm$29  & 4.52$\pm$0.15 & 1.54$\pm$0.07 & 0.06$\pm$0.04    & 38,8  & 0.03,0.06 & [a]  & 1.17 & 4.41\\
\object{HD\,219449} & 4757$\pm$102 & 2.71$\pm$0.25 & 1.71$\pm$0.09 & 0.05$\pm$0.14    & 31,8  & 0.12,0.09 & [b]  &  --  &   --\\
\object{HD\,330075} & 5017$\pm$53  & 4.22$\pm$0.11 & 0.69$\pm$0.08 & 0.08$\pm$0.06    & 35,5  & 0.05,0.20 & [f]  & 0.69 & 4.37\\
\hline
\end{tabular}
\\ $\dagger$ The instruments used to obtain the spectra were: [a] 2.2-m ESO/FEROS; [b] TNG/SARG; [c] From Paper\,IV
\citep[][]{Santos-2004b}; [d] VLT Kueyen/UVES; [e-i] HARPS -- From \citet[][]{Santos-2004a}, \citet[][]{Pepe-2004}, 
\citet[][]{Moutou-2005}, \citet[][]{Lovis-2005} and \citet[][]{Eggenberger-2005}, respectively (stellar parameters
and [Fe/H] derived by our team).
\label{tab:planets}
\end{table*}

\begin{table*}[th!]
\caption[]{Comparison sample stars presented in this paper. See text for more details.}
\begin{tabular}{lcccrccccc}
\hline
HD     & T$_{\mathrm{eff}}$ & $\log{g}_{spec}$ & $\xi_{\mathrm{t}}$ & \multicolumn{1}{c}{[Fe/H]} & N(\ion{Fe}{i},\ion{Fe}{ii}) & $\sigma$(\ion{Fe}{i},\ion{Fe}{ii}) & Instr.$^a$ & Mass         & $\log{g}_{hipp}$ \\
number & [K]                &  [cm\,s$^{-2}$]  &  [km\,s$^{-1}$]  &        &                              &                                    &            & [M$_{\sun}$] & [cm\,s$^{-2}$] \\
\hline
\object{HD\,2151  }  & 5837$\pm$30     & 4.00$\pm$0.12 & 1.36$\pm$0.05 & $-$0.08$\pm$0.04 & 36,6   & 0.03,0.05& [1]& 1.24    & 4.01\\
\object{HD\,4747  }  & 5316$\pm$38     & 4.48$\pm$0.10 & 0.79$\pm$0.06 & $-$0.21$\pm$0.05 & 36,6   & 0.04,0.05& [1]& 0.79    & 4.54\\
\object{HD\,21175 }  & 5274$\pm$58     & 4.46$\pm$0.15 & 0.86$\pm$0.07 &  0.16$\pm$0.06 & 35,6   & 0.05,0.08& [1]& 0.88    & 4.55\\
\object{HD\,32147 }  & 4705$\pm$75     & 4.12$\pm$0.31 & 0.60$\pm$0.17 &  0.30$\pm$0.08 & 35,4   & 0.07,0.15& [1]& --	   & --  \\
\object{HD\,52698 }  & 5177$\pm$51     & 4.44$\pm$0.09 & 0.98$\pm$0.06 &  0.16$\pm$0.06 & 37,6   & 0.05,0.03& [1]& 0.83    & 4.54\\
\object{HD\,52919 }  & 4740$\pm$49     & 4.25$\pm$0.15 & 1.03$\pm$0.09 & $-$0.05$\pm$0.06 & 37,6   & 0.07,0.06& [2]& 0.76    & 4.72\\
\object{HD\,53143 }  & 5462$\pm$54     & 4.47$\pm$0.16 & 1.08$\pm$0.07 &  0.22$\pm$0.06 & 39,8   & 0.06,0.06& [2]& 1.00    & 4.59\\
\object{HD\,57095 }  & 4945$\pm$54     & 4.45$\pm$0.29 & 1.09$\pm$0.08 & $-$0.03$\pm$0.06 & 38,9   & 0.06,0.15& [2]& 0.62    & 4.29\\
\object{HD\,61606 }  & 4958$\pm$80     & 4.45$\pm$0.19 & 1.01$\pm$0.12 &  0.01$\pm$0.08 & 38,9   & 0.07,0.07& [2]& 0.78    & 4.62\\
\object{HD\,64606 }  & 5351$\pm$68     & 4.62$\pm$0.42 & 0.92$\pm$0.19 & $-$0.71$\pm$0.09 & 39,10  & 0.07,0.19& [2]& 0.69    & 4.59\\
\object{HD\,65277 }  & 4843$\pm$80     & 4.39$\pm$0.58 & 0.99$\pm$0.13 & $-$0.26$\pm$0.08 & 37,9   & 0.07,0.31& [2]& 0.74    & 4.70\\
\object{HD\,65486 }  & 4660$\pm$66     & 4.55$\pm$0.21 & 0.82$\pm$0.16 & $-$0.33$\pm$0.07 & 38,6	& 0.08,0.10& [2]& 0.61    & 4.61\\
\object{HD\,67199 }  & 5136$\pm$56     & 4.54$\pm$0.13 & 0.84$\pm$0.08 &  0.06$\pm$0.06 & 38,9	& 0.05,0.05& [2]& 0.80    & 4.55\\
\object{HD\,82342 }  & 4907$\pm$78     & 4.46$\pm$0.51 & 0.74$\pm$0.15 & $-$0.40$\pm$0.08 & 35,5	& 0.07,0.25& [2]& 0.74    & 4.75\\
\object{HD\,85512 }  & 4505$\pm$176    & 4.71$\pm$0.96 & 0.32$\pm$1.15 & $-$0.18$\pm$0.19 & 41,4	& 0.17,0.54& [2]& 0.63    & 4.65\\
\object{HD\,100623}  & 5246$\pm$37     & 4.54$\pm$0.23 & 1.00$\pm$0.06 & $-$0.38$\pm$0.05 & 39,11  & 0.04,0.11& [2]& 0.75    & 4.60\\
\object{HD\,101581}  & 4646$\pm$96     & 4.80$\pm$0.39 & 0.58$\pm$0.35 & $-$0.37$\pm$0.09 & 37,6	& 0.09,0.20& [2]& 0.74    & 4.75\\
\object{HD\,102365}  & 5667$\pm$27     & 4.59$\pm$0.14 & 1.05$\pm$0.06 & $-$0.27$\pm$0.04 & 39,11  & 0.03,0.06& [2]& 0.85    & 4.43\\
\object{HD\,102438}  & 5639$\pm$51     & 4.60$\pm$0.16 & 1.12$\pm$0.10 & $-$0.23$\pm$0.06 & 39,12  & 0.06,0.07& [2]& 0.88    & 4.50\\
\object{HD\,103932}  & 4874$\pm$125    & 4.44$\pm$0.69 & 1.30$\pm$0.17 &  0.09$\pm$0.13 & 39,7	& 0.12,0.37& [2]& 0.84    & 4.81\\
\object{HD\,104304}  & 5562$\pm$50     & 4.37$\pm$0.14 & 1.10$\pm$0.06 &  0.27$\pm$0.06 & 39,10  & 0.05,0.06& [2]& 0.93    & 4.40\\
\object{HD\,109200}  & 5103$\pm$46     & 4.47$\pm$0.23 & 0.75$\pm$0.07 & $-$0.24$\pm$0.05 & 37,10  & 0.05,0.11& [2]& 0.70    & 4.51\\
\object{HD\,111261}  & 4529$\pm$62     & 4.44$\pm$0.64 & 0.78$\pm$0.17 & $-$0.35$\pm$0.08 & 34,5	& 0.08,0.37& [2]& 0.51    & 4.57\\
\object{HD\,115617}  & 5577$\pm$33     & 4.34$\pm$0.11 & 1.07$\pm$0.04 &  0.01$\pm$0.05 & 39,11  & 0.04,0.05& [2]& 0.90    & 4.43\\
\object{HD\,118972}  & 5241$\pm$66     & 4.43$\pm$0.42 & 1.24$\pm$0.08 & $-$0.01$\pm$0.08 & 39,11  & 0.07,0.20& [2]& 0.90    & 4.63\\
\object{HD\,125072}  & 5001$\pm$115    & 4.39$\pm$0.52 & 1.21$\pm$0.15 &  0.24$\pm$0.11 & 38,10  & 0.11,0.27& [2]& 0.84    & 4.63\\
\object{HD\,128620}  & 5844$\pm$42     & 4.30$\pm$0.19 & 1.18$\pm$0.05 &  0.28$\pm$0.06 & 39,12  & 0.05,0.08& [2]& 1.09    & 4.32\\
\object{HD\,128621}  & 5199$\pm$80     & 4.37$\pm$0.27 & 1.05$\pm$0.10 &  0.19$\pm$0.09 & 39,9	& 0.08,0.12& [2]& 0.81    & 4.47\\
\object{HD\,130992}  & 4751$\pm$73     & 4.32$\pm$0.29 & 0.72$\pm$0.13 & $-$0.06$\pm$0.07 & 37,6	& 0.07,0.14& [2]& 0.40    & 4.30\\
\object{HD\,131977}  & 4693$\pm$80     & 4.36$\pm$0.25 & 0.97$\pm$0.16 &  0.07$\pm$0.10 & 38,8	& 0.09,0.11& [2]& 0.55    & 4.49\\
\object{HD\,135204}  & 5332$\pm$37     & 4.31$\pm$0.13 & 0.84$\pm$0.05 & $-$0.11$\pm$0.05 & 38,10  & 0.04,0.06& [2]& 0.76    & 4.37\\
\object{HD\,136352}  & 5667$\pm$43     & 4.39$\pm$0.07 & 1.08$\pm$0.09 & $-$0.31$\pm$0.06 & 38,10  & 0.05,0.03& [2]& 0.83    & 4.33\\
\object{HD\,140901}  & 5645$\pm$37     & 4.40$\pm$0.25 & 1.14$\pm$0.05 &  0.13$\pm$0.05 & 39,11  & 0.04,0.11& [2]& 0.99    & 4.50\\
\object{HD\,142709}  & 4806$\pm$106    & 4.60$\pm$0.59 & 1.09$\pm$0.19 & $-$0.17$\pm$0.12 & 38,6	& 0.11,0.31& [2]& 0.78    & 4.85\\
\object{HD\,144628}  & 5071$\pm$43     & 4.41$\pm$0.19 & 0.82$\pm$0.07 & $-$0.36$\pm$0.06 & 38,8	& 0.05,0.08& [2]& 0.70    & 4.59\\
\object{HD\,145417}  & 5045$\pm$111    & 4.74$\pm$0.36 & 0.90$\pm$0.43 & $-$1.10$\pm$0.15 & 34,4	& 0.13,0.16& [2]& 0.62    & 4.73\\
\object{HD\,146233}  & 5786$\pm$35     & 4.31$\pm$0.19 & 1.18$\pm$0.05 &  0.08$\pm$0.05 & 38,12  & 0.04,0.08& [2]& 1.01    & 4.43\\
\object{HD\,149661}  & 5290$\pm$52     & 4.39$\pm$0.21 & 1.11$\pm$0.07 &  0.04$\pm$0.06 & 39,11  & 0.06,0.10& [2]& 0.90    & 4.60\\
\object{HD\,150689}  & 4867$\pm$87     & 4.43$\pm$0.34 & 1.12$\pm$0.13 & $-$0.08$\pm$0.08 & 36,7	& 0.07,0.16& [2]& 0.78    & 4.69\\
\object{HD\,152391}  & 5521$\pm$43     & 4.54$\pm$0.24 & 1.29$\pm$0.06 &  0.03$\pm$0.05 & 38,12  & 0.04,0.12& [2]& 1.04    & 4.64\\
\object{HD\,154088}  & 5414$\pm$60     & 4.28$\pm$0.24 & 1.14$\pm$0.07 &  0.33$\pm$0.07 & 39,11  & 0.07,0.11& [2]& 0.93    & 4.46\\
\object{HD\,154577}  & 4973$\pm$55     & 4.73$\pm$0.14 & 0.94$\pm$0.12 & $-$0.63$\pm$0.07 & 36,7	& 0.05,0.06& [2]& 0.68    & 4.68\\
\object{HD\,156026}  & 4568$\pm$94     & 4.67$\pm$0.76 & 0.60$\pm$0.26 & $-$0.18$\pm$0.09 & 36,5	& 0.09,0.44& [2]& 0.70    & 4.75\\
\object{HD\,156274}  & 5300$\pm$32     & 4.41$\pm$0.19 & 1.00$\pm$0.05 & $-$0.33$\pm$0.04 & 39,9	& 0.04,0.09& [2]& 0.72    & 4.48\\
\object{HD\,156384}  & 4819$\pm$110    & 4.66$\pm$1.02 & 1.04$\pm$0.22 & $-$0.43$\pm$0.12 & 39,6	& 0.11,0.53& [2]& 0.45    & 4.41\\
\object{HD\,165185}  & 5942$\pm$85     & 4.53$\pm$0.13 & 1.39$\pm$0.16 &  0.02$\pm$0.10 & 35,10  & 0.08,0.05& [2]& 1.12    & 4.52\\
\object{HD\,165499}  & 5950$\pm$45     & 4.31$\pm$0.13 & 1.14$\pm$0.08 &  0.01$\pm$0.06 & 39,11  & 0.05,0.05& [2]& 1.06    & 4.30\\
\object{HD\,170493}  & 4854$\pm$120    & 4.49$\pm$0.43 & 1.36$\pm$0.20 &  0.15$\pm$0.13 & 39,8	& 0.12,0.21& [2]& 0.81    & 4.68\\
\object{HD\,170573}  & 4767$\pm$89     & 4.44$\pm$0.5  & 1.07$\pm$0.14 & $-$0.07$\pm$0.09 & 38,6	& 0.09,0.27& [2]& 0.77    & 4.78\\
\object{HD\,170657}  & 5115$\pm$52     & 4.48$\pm$0.28 & 1.13$\pm$0.07 & $-$0.22$\pm$0.06 & 39,9	& 0.06,0.14& [2]& 0.77    & 4.60\\
\object{HD\,172051}  & 5634$\pm$30     & 4.43$\pm$0.14 & 1.06$\pm$0.05 & $-$0.21$\pm$0.04 & 39,11  & 0.04,0.05& [2]& 0.90    & 4.53\\
\object{HD\,177565}  & 5664$\pm$28     & 4.43$\pm$0.09 & 1.02$\pm$0.04 &  0.14$\pm$0.04 & 38,12  & 0.03,0.05& [2]& 0.98    & 4.46\\
\object{HD\,190248}  & 5614$\pm$34     & 4.29$\pm$0.07 & 1.10$\pm$0.04 &  0.36$\pm$0.04 & 38,8	& 0.04,0.04& [1]& 1.02    & 4.31\\
\hline
\end{tabular}
\\ $^a$ The instruments used to obtain the spectra were: [1] 1.2-m Swiss Telescope/CORALIE; [2] 2.2-m ESO/FEROS
\label{tab:comparison2}
\end{table*}

Interestingly, these observational results are now being corroborated by
theoretical models of planet formation, and give support to the core-accretion
model to explain the origin of the now discovered giant
planets \citep[][]{Pollack-1996,Alibert-2004b}. Current models can even reasonably
explain the observed [Fe/H] distribution of planet-host stars \citep[][]{Ida-2004a,Ida-2004b}. 
However, and as discussed in \citet[][]{Santos-2004b}, 
current data still cannot exclude that giant planets may also be formed by the 
disk instability model \citep[][]{Boss-1997,Boss-2002,Mayer-2002}, 
in particular planets orbiting low metallicity stars.

In the last few years we have published a series of papers regarding the study of
the stellar metallicity-giant planet connection 
\citep[][]{Santos-2000b,Santos-2001,Santos-2003,Santos-2004b} 
(hereafter Papers I, II, III and IV, respectively). In the last three of
these papers we have presented a spectroscopic analysis of a sample of 41 ``single'' 
field stars, used as a reference to study the excess metallicity observed in
planet hosts. In Paper\,IV we presented a uniform
spectroscopic analysis of 98 extra-solar planet-host stars, including most
of the known objects.
The continuation of this work is extremely important, since the
search for new possible correlations between the presence (and properties) 
of exoplanets and the chemical abundances of their host stars needs increased samples. 

In this paper we add 53 new stars to our ``comparison''
sample, as well as precise and uniform metallicity estimates for a further 29 
known planet-hosts, most of which have no previous spectroscopic metallicity estimate. 
In Sects.\,\ref{sec:samples} and \ref{sec:analysis} 
we present the observed samples, and in Sect.\,\ref{sec:distributions} we review the metallicity 
distribution of planet-host stars, as well as of the whole volume-limited comparison 
sample of 94 solar-type F-G-K dwarfs. The analysis confirms, as expected, that stars 
with planets are clearly metal-rich when compared to average field dwarfs. 
In Sect.\,\ref{sec:bias} we review a few possible biases that should be 
considered when studying the metallicity of planet-host stars. 
In Sect.\,\ref{sec:pib} we compare the metallicity distributions for
``single'' planet-host stars with that found for planet-hosts in
multiple stellar systems. We conclude in Sect.\,\ref{sec:conclusions}.

\section{The samples}
\label{sec:samples}

\subsection{New comparison sample}

In Papers\,2 through 4 we compared the metallicity distribution for stars
with planets with a volume-limited sample of stars
having right-ascensions (RA) between 20$^h$ and 9$^h$, and a 
distance below 20\,pc as derived from Hipparcos parallaxes \citep[][]{ESA-1997}. 
All these stars belong to the CORALIE southern planet search sample \citep[][]{Udry-2000},
have spectral types between F8 and M1, 
and were not found to harbor any planetary companion.

In order to extend the number of comparison stars in our samples, we have now 
observed a complementary sample of CORALIE stars with RA between 
9$^h$ and 20$^h$ (and d$<$20\,pc).
These stars were observed using the FEROS spectrograph at the 2.2-m ESO/MPI
telescope (ESO, La Silla) in March 2004 (observing run ID 072.C-0033).
Data reduction was done using the FEROS pipeline, and the wavelength calibration 
was done using the spectrum of a ThAr lamp. The final spectra have a resolution 
R=$\Delta\lambda/\lambda$=48\,000, and a the S/N between $\sim$200 and 400.

A few stars from the former comparison sample for which we could not obtain 
a spectrum before were also observed, both using FEROS and the CORALIE spectrograph at the 
1.2-m Euler Swiss telescope, also at La Silla observatory (R=50\,000). 

The comparison sample presented in this paper is listed in Table\,\ref{tab:comparison2}. 
Together with Table\,5 of Paper\,IV, this constitutes a large volume-limited sample of 94 ``single'' stars (with no known planetary-mass companions).
Amid the stars in this new volume-limited comparison sample there is \object{HD\,147513}
and the very metal rich planet-host star \object{HD\,160691} \citep[][]{Butler-2001,Santos-2004a}. 
For the list of other planet-hosts that belong to the whole volume-limited comparison 
sample we refer to \citet[][]{Santos-2004b}.

\subsection{Planet-host stars}

In addition to the new comparison sample, spectra of some planet hosts was also obtained,
both during the March run and in two complementary runs, one in October 2004 (using FEROS -- 
observing run ID 074.C-0135) and another in December 2004, using the UVES spectrograph
at the VLT 8.2-m Kueyen telescope (run ID 074.C-0134). These latter spectra were reduced using
the UVES pipeline and have a resolution R$\sim$75\,000. 

For \object{HD\,104985} and \object{HD\,219449} we further obtained spectra 
with a resolution of 57\,000 using the SARG spectrograph, at the TNG telescope 
(La Palma Observatory, Spain). For a few other recently-announced planet-hosts,
stellar parameters were derived using the HARPS spectrograph \citep[R$\sim$100\,000 -- for
details see e.g.][]{Pepe-2004}. These parameters were also presented in
the papers announcing the discovery of the exoplanets (see Table\,\ref{tab:planets}).

The list of planet-host stars presented in this paper 
is shown in Table\,\ref{tab:planets}. For planet-discovery references see the 
Geneva Planet-Search web pages\footnote{http://obswww.unige.ch/exoplanets}.

Adding the 97 planet-host
stars whose [Fe/H] values were listed in Paper\,IV\footnote{\object{HD\,219542\,B} 
was excluded from the list of planet hosts presented 
in Paper IV, as its planet has been denied \citep[][]{Desidera-2004}.} to the 
results listed in Table\,\ref{tab:planets}, we now have
a uniform sample of metallicities and stellar parameter for 119 stars
with orbiting planets. From the whole known planet-host star sample, the 
only 4 objects for which we could not obtain metallicity estimates are \object{GJ\,876} 
and \object{GJ\,436} \citep[both M-dwarfs -- see][]{Bonfils-2005}, \object{HD\,45350} and \object{HD\,13189} (a giant star). 
To these we add the transiting planets detected by the OGLE survey 
\citep[][]{Konacki-2003,Bouchy-2004,Pont-2004}, three currently with poor estimates given 
the low quality of the now available spectra, as well as 
\object{TrEs-1} \citep[][]{Alonso-2004}.

\section{Spectroscopic analysis}
\label{sec:analysis}

The spectroscopic analysis was done in LTE using the 
2002 version of the code MOOG \citep[][]{Sneden-1973}\footnote{The source code of MOOG2002 can be 
downloaded at http://verdi.as.utexas.edu/moog.html} and a grid of Kurucz Atlas plane-parallel 
model atmospheres \citep[][]{Kurucz-1993}. 

Stellar parameters and metallicities were derived as in \citet[][]{Santos-2004b}, 
based on the Equivalent Widths (EW) of 39 \ion{Fe}{i} and 12 \ion{Fe}{ii} week lines, and by imposing 
excitation and ionization equilibrium. The line EW were measured using IRAF
\footnote{IRAF is distributed by National Optical Astronomy Observatories, operated
by the Association of Universities for Research in Astronomy, Inc.,
under contract with the National Science Foundation, U.S.A.} ``splot'' and ``bplot'' 
routines within the {\tt echelle} package. 
A Fortran\,77 code that uses a Downhill Simplex Method 
\citep[][]{Numerical-1992} was then used to find the best solution in 
the (stellar) parameter space once the line EWs are measured. 

The results of this analysis are presented in Tables\,\ref{tab:planets} and \ref{tab:comparison2}.
As shown in Paper\,IV, this methodology gives excellent results for the derived stellar
parameters, compatible with other spectroscopic results.
A comparison between the EW
and stellar parameters derived from these spectra and data obtained using
other instruments also shows that no significant differences 
exist \citep[e.g.][]{Santos-2004b} -- see also Table\,\ref{tab:planets}.

The errors in T$_{\mathrm{eff}}$, $\log{g}$, $\xi_t$ and [Fe/H] were derived as
in Paper\,IV, and are typically of
the order of 50\,K, 0.2\,dex, 0.1\,km\,s$^{-1}$
and 0.05\,dex, respectively. Higher uncertainties are found for 
lower temperature stars.

In the Tables we also present the stellar masses. These were 
derived from an interpolation of the theoretical isochrones of \citet[][]{Schaller-1992},
\citet[][]{Schaerer-1993a} and \citet[][]{Schaerer-1993b}, using $M_{V}$ computed using Hipparcos
parallaxes \citep[][]{ESA-1997}, a Bolometric Correction (BC) from \citet[][]{Flower-1996}
and the T$_{\mathrm{eff}}$ obtained from the spectroscopy. Although this method
to estimate individual stellar masses may suffer from important 
uncertainties \citep[][]{Fernandes-2004}, from a statistical point of view it 
should provide good results.
We adopt a typical relative error of 0.05\,M$_{\sun}$ for the masses. In some cases,
no mass estimates are presented, since these involved large extrapolations of the
isochrones.

From the derived stellar mass, the spectroscopic temperature and BC, we could also 
obtain ``trigonometric'' surface gravities for our stars using 
the basic formula presented in Paper\,IV. The results (see Tables\,\ref{tab:planets} 
and \ref{tab:comparison2}) show that the two gravity estimates are perfectly compatible 
within the errors, as expected from our previous studies (see Paper\,IV). Considering also
all planet-hosts and comparison sample stars presented in Paper\,IV, the average
difference between the two estimates is nearly zero (0.004\,dex), with a stardard 
deviation of 0.14\,dex.

The stellar parameters for \object{HD\,41004\,A} may suffer from
the fact that the spectrum of this star is blended with the spectrum of the
fainter M-dwarf companion \object{HD\,41004\,B} \citep[see e.g. ][]{Santos-2002a}.
The separation between the two components of the binary is only 0.5'', smaller than the 
diameter of the FEROS fiber. However, given the magnitude difference between the 
two components, the flux of \object{HD\,41004\,B} should represent a mere 3\% 
of the total flux of the primary; the impact on the determination of the line 
EW for \object{HD\,41004\,A} is thus probably limited.

\subsection{Temperature scale}

In a recent paper, \citet[][]{Ramirez-2004} derived temperatures and radii for
extra-solar planet-host stars using the IRFM technique. In their work
they compare the temperatures derived in Paper\,IV, as well
as by \citet[][]{Ribas-2003} (who used infra-red photometry), with the ones obtained by them, 
and conclude that our values are systematically shifted by $\sim$100\,K. 

One argument used by \citet[][]{Ramirez-2004} in favor of their temperature
scale is the fact that for 5 stars that have accurate stellar radii measured
\citep[and thus accurate temperatures ][]{Ramirez-2004}, the results
from the IRFM differ by only 14\,K with respect to the measured ``direct'' 
temperatures (see their Table\,1). 

Four of the stars in their Table\,1 were also analyzed by us. 
These are \object{HD\,128620}, \object{HD\,128621} (see Table\,\ref{tab:comparison2}), 
\object{HD\,10700} and \object{HD\,209458} (see Paper\,IV). Interestingly, a comparison of 
the obtained T$_{\mathrm{eff}}$ shows that the difference between our estimates and the ``direct''
temperatures listed by \citet[][]{Ramirez-2004} is only of 7\,K. In other words,
we do not find any strong reason to support that the IRFM scale of temperatures 
is more accurate than the one derived from the method used in our studies 
(Paper\,IV and this paper).

\begin{figure}[t]
\psfig{width=\hsize,file=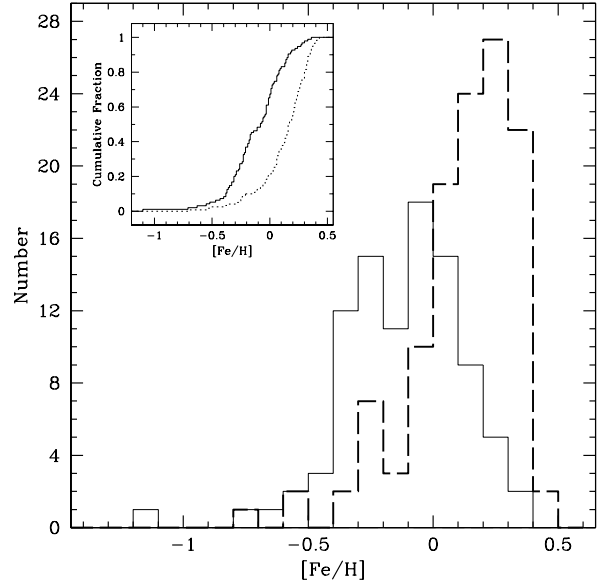}
\caption[]{Metallicity distribution for 119 planet-host stars (dashed line) and for
a volume-limited ``comparison'' sample of 94 stars with no known planetary-mass 
companions (continuous line). The metallicities were taken from Table\,\ref{tab:planets}, 
Table\,\ref{tab:comparison2} and from Tables\,2-5 of \citet[][]{Santos-2004b}. The difference between 
the average metallicity of two samples is 0.24\,dex. {\it Inset}: cumulative functions
for both distributions. A Kolmogorov-Smirnov test shows that the probability that
both distributions belong to the same population is of the order of 10$^{-12}$.}
\label{fig:histo}
\end{figure}

The major goal of the current study is to compare 
the [Fe/H] values for planet-host stars and ``single'' field dwarfs in an 
uniform way. The crucial point is thus to use uniform samples, 
analyzed using the same line-lists, model atmospheres and techniques 
to derive stellar parameters and metallicities. This goal is already achieved, 
and thus our conclusions are not dependent on the temperature scale used. The definition 
of a correct scale, which is indeed very important when an absolute study is needed,
is thus not a limitation in our case.

%

\section{Metallicity distributions}
\label{sec:distributions}

In Fig.\,\ref{fig:histo} we present the metallicity distribution for
the 119 planet-host stars with metallicity estimates (Paper\,IV and Table\,\ref{tab:planets}), 
as well as for the whole volume-limited comparison sample of 94 ``single'' 
dwarfs (Paper\,IV and Table\,\ref{tab:comparison2}). 

As it is well seen from the plot, and confirming previous studies, planet hosts
have, on average, a [Fe/H] that is 0.24\,dex higher than that found for
stars with no detected planetary mass companions. 
The Kolmogorov-Smirnov probability that the two samples belong 
to the same population is of the order of 10$^{-12}$.

Although it is not our goal here to study in detail the metallicity
distribution of planet-hosts (see next Section), or to study the frequency of planets
as a function of stellar metallicity (see Paper\,IV), 
the addition of new stars to our comparison sample reinforces the 
results presented in previous works (e.g. Papers II and IV).
For instance, the average [Fe/H] of the
stars in Table\,\ref{tab:comparison2} is $-$0.11, similar to the one
found for the 41 comparison stars presented in Paper\,IV. 

The metallicity distribution for our 94 comparison sample stars though symetrical, seems
to have a small depression for stars with [Fe/H] values
between $-$0.1 and $-$0.2 dex. A similar scarcity of planet-host
stars seems to exist at the same [Fe/H] interval. The cause for this
feature, also tentatively observed in our previous samples 
\citep[for iron and for other elements --][]{Santos-2001,Bodaghee-2003} as well
as in the volume-limited sample of \citet[][]{Fischer-2005}, is not 
clear, but may be simply related to small number statistics.
It is interesting to note that this metallicity regime is at the 
frontier of the Galactic thin and thick disk populations \citep[e.g.][]{Fuhrmann-2004}.
Further studies of the UVW velocities of the stars in our sample
as well as of other chemical elements are necessary to better
tackle this issue.

\subsection{Sampling biases}
\label{sec:bias}

In Paper\,III we have shown that the whole metallicity ``excess'' observed for
stars with giant planets cannot be explained by any observational or
sampling bias. However, the striking fact that stars with planets are
particularly metal-rich has led some planet-search programs to base their samples 
on metal-rich stars \citep[e.g.][]{Tinney-2002,Fischer-2004}. 
Such surveys, now giving their first results, are adding to the lists of known
planet hosts some objects that should be considered with care if
a consistent statistical study about the stellar metallicity-giant
planet connection is pursued.

The impact of these surveys may have consequences on any future 
statistical studies of the properties of exoplanets orbiting
stars with different chemical composition. For example, the first
result of these metallicity-biased surveys will be the discovery
of short period giant planets (hot-Jupiters). Any
study of the relation between the metallicity of the stars
and the orbital period of the orbiting planet 
\citep[][]{Gonzalez-1998,Queloz-2000,Santos-2003,Sozzetti-2004} will thus be
seriously compromised, as such systematic errors are
very difficult to take into account.

Such was not taken in consideration when plotting stars in Fig.\,\ref{fig:histo},
as the goal of the present paper is simply to present stellar metallicities
for an extended comparison sample, and for a few planet-host stars
whose stellar parameters and [Fe/H] were not presented in Paper\,IV. 
A detailed and unbiased comparison of the [Fe/H] distributions 
for planet-hosts and ``single'' stars implies the use of samples for which we are sure
not to have any important bias. This was the case of our
previous studies (e.g. Paper\,IV), where we have compared planet-hosts found amid
the stars in the CORALIE planet-search sample \citep[][]{Udry-2000}
with the [Fe/H] distribution of the whole sample itself; with CORALIE, a target 
was never chosen based on its high metal content.

\section{Planets in multiple stellar systems}
\label{sec:pib}

\begin{figure}[t]
\psfig{width=\hsize,file=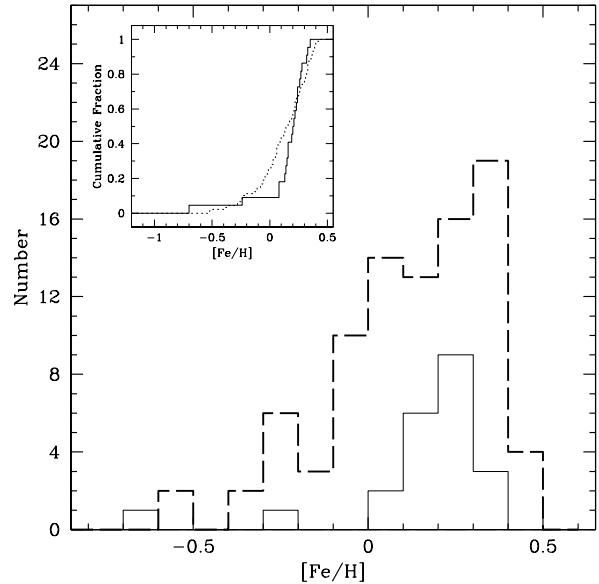}
\caption[]{Metallicity distribution for 22 planet-host stars in binary stellar systems (PIB -- filled line) when
compared with the same distribution for single planet-hosts (dashed line). 
The metallicities were taken from Table\,\ref{tab:planets} and from \citet[][]{Santos-2004b}. 
The difference between the average metallicity of two samples is $\sim$0.03\,dex, PIB stars being slighly overabundant. {\it Inset}: cumulative functions
for both distributions. A Kolmogorov-Smirnov test shows that the probability that
both distributions belong to the same population is of the order of 0.08.}
\label{fig:pib}
\end{figure}

In Fig.\,\ref{fig:pib} we present a comparison between the [Fe/H]
distributions of 22 stars with planets that are currently known to belong to multiple stellar 
systems\footnote{Namely 
\object{HD\,142},
\object{HD\,9826},
\object{HD\,13445},
\object{HD\,16141},
\object{HD\,19994},
\object{HD\,40979},
\object{HD\,41004\,A},
\object{HD\,46375},
\object{HD\,75289},
\object{HD\,75732},
\object{HD\,80606},
\object{HD\,89744},
\object{HD\,99492},
\object{HD\,114762},
\object{HD\,120136},
\object{HD\,142022\,A},
\object{HD\,178911\,B},
\object{HD\,186427},
\object{HD\,190360},
\object{HD\,195019},
\object{HD\,222404} and
\object{BD\,-10\,3166}.} -- mostly binary systems, hereafter PIB -- 
with the same distribution for ``single'' planet-host stars. See \citet[][]{Eggenberger-2004} and Eggenberger et al. (in prep.) for references.

The comparison shows that no major difference seems to exist between the two
samples. Even if PIB stars seem to
be slighly over-metallic when compared to single planet-hosts, the Kolmogorov-Smirnov 
probability that the two samples belong to the same population is 0.08. If we 
exclude \object{HD\,114762} from the PIB sample \citep[this star has
a companion with a minimum mass above 10 times the mass of Jupiter, likely a 
brown dwarf][]{Latham-1989}, this probability decreases to 0.04, 
only marginally significant. 

It has been proposed that planet migration processes may act differently in single
and multiple stellar systems, and that migration rates may be increased if
the planet-host has a stellar companion \citep[for a discussion see][]{Eggenberger-2004}. 
Increasing the migration efficiency could then increase the scattering within the
disc of planetesimals \citep[e.g.][]{Murray-1998}. In such a case, we could 
expect that PIB stars, or at least those with ``close'' stellar companions, could have been
polluted by a larger amount of planetesimal (metal-rich) material. 

The fact that we do not find a clear difference between PIB stars and single
planet-hosts suggests that strong planetesimal accretion is not usual both
for single planet-hosts \citep[][]{Pinsonneault-2001,Santos-2003,Cody-2004}
nor for planet-host stars in multiple stellar systems.
However, if the slight correlation discussed above is real it
may imply, for example, that stellar pollution played some role (although probably 
small) in planet-hosts in multiple stellar systems. A follow-up of these results as more planets 
are discovered is necessary to resolve this issue.

We also checked if within the PIB star sample there was any correlation between
stellar metallicity and the separation of the stellar components. No major correlation
was found.

\section{Concluding remarks}
\label{sec:conclusions}

In this paper we have presented accurate stellar metallicities for
a volume-limited sample of 53 comparison ``single'' stars. This sample
was observed to complement the already existing list of 41 comparison 
dwarfs presented in Papers\,II and IV.

We further present metallicities and stellar parameters for 29 planet-host stars, 
most of which were not studied in our previous series of papers (Papers\,I-IV).
Together with the 97 planet host stars studied in Paper\,IV, we now have
a large sample of 119 stars with planets (almost all the known targets) 
with precise metallicity and stellar parameter estimates. 

These two large samples (of planet-hosts and ``single'' stars) were
analyzed using the same line-lists, model atmospheres and
techniques to derive stellar parameters and [Fe/H]. Therefore, they
constitute very uniform samples, unique and appropriate for future studies
of the stellar metallicity-giant planet connection.

We have used these samples to investigate whether stars with planets
belonging to multiple stellar systems have a different [Fe/H]
distribution when compared with single planet-hosts. The results
show that a slight tendency, although not statistically
significant, may exist. The enlargement of the samples is needed to
resolve this issue.

\begin{acknowledgements}
  We would like to thank the anonymous referee for a prompt
  report. Support from Funda\c{c}\~ao para a Ci\^encia e Tecnologia (Portugal) 
  to N.C.S. in the form of a scholarship is gratefully acknowledged.
  We wish to thank the Swiss National Science Foundation (Swiss NSF) 
  for the continuous support for this project.
\end{acknowledgements}

\bibliographystyle{aa}
\bibliography{santos_bibliography}

\end{document}